# Anomalous Nonlinear Microwave Response of Epitaxial YBa$_2$Cu$_3$O$_{7-x}$ Films on MgO

M. A. Hein, Dept. of Physics, University of Wuppertal, Germany.

P. J. Hirst and R. G. Humphreys, DERA Malvern, Worcestershire, UK.

D. E. Oates, MIT Lincoln Laboratory, Lexington/MA, U.S.A.

A. V. Velichko, School of Electronic and Electrical Engineering, University of Birmingham, UK.

Abstract − We have investigated the anomalous nonlinear microwave response of epitaxial electron-beam coevaporated YBa$_2$Cu$_3$O$_{7-\delta}$ films on MgO. The power and temperature dependent surface impedance and two-tone intermodulation distortion were measured in stripline resonators at several frequencies between 2.3 and 11.2 GHz and at temperatures between 1.7 K and $T_c$. All of the seven films measured to date show a decrease of the surface resistance $R_s$ of up to one order of magnitude as the microwave current is increased up to approximately 1 mA for $T < 20$ K. The surface reactance $X_s$ showed only a weak increase in the same region. The usual nonlinear increase of $R_s$ and $X_s$ was observed at high currents in the 100-mA range. The minimum of $R_s$ correlates with a pronounced plateau in the third-order intermodulation signal. We have developed a phenomenological two-fluid model, incorporating a microwave current-dependent quasiparticle scattering rate $g$ and normal fraction $f_n$. The model contains only three adjustable parameters. We find good agreement with the measured data for constant $f_n$ and a current-dependent scattering rate $g$ whose magnitude increases almost linearly with temperature and frequency. Obvious mechanisms leading to nonlinear microwave response like Josephson coupling across grain boundaries, nonlocal or non-equilibrium effects cannot explain the data. The anomalies could rather reflect the specific charge and spin ordering and transport phenomena in the low-dimensional cuprate superconductors.

**Corresponding Author:**

Dr. Matthias Hein,

Dept. of Physics, University of Wuppertal, Gauss-Strasse 20, D-42097 Wuppertal, Germany

Tel. +49 (202) 439-2747

Fax +49 (202) 439-2811

Email: mhein@venus.physik.uni-wuppertal.de

# I. INTRODUCTION

Understanding and controlling the nonlinear microwave properties of high-temperature superconductors (HTS) is important for material and device engineering. On the material side, comprehensive insight into charge and spin ordering and transport in the low-dimensional cuprates cannot be attained without detailed knowledge of dynamic processes like pair breaking and quasi-particle scattering [1]. Furthermore, studies of the microwave response provide essential information on phase purity, grain connectivity and interface effects, which cannot readily be obtained at similar resolution by other techniques [2,3]. This information is important in order to optimise materials like epitaxial films or biaxially textured tapes with respect to critical current density, surface impedance, and power handling. On the device side, the nonlinear microwave response of HTS is relevant for applications of very narrow–band receive filters, which are susceptible to intermodulation distortion, or high-power transmit filters, which suffer from power-dependent absorption [4-6].

Despite numerous attempts worldwide, the sources of nonlinear microwave response are not understood at present, mainly because of the variety of techniques employed for the preparation and characterisation of samples and of a lack of systematic studies. While most reports of nonlinear surface impedance $Z_s$ indicate conventional behaviour, several groups have recently reported unexpected microwave properties like the recovery of superconductivity in DC or microwave magnetic fields [7-11] or marked deviations of the third-order intermodulation product or harmonic generation from the expected cubic dependence on input power [12,13]. These frequently observed effects, which are now commonly summarised as "anomalous" behaviour, challenge our experimental and theoretical approaches to understand their physical origin.

The majority of previous research projects focused on the surface impedance of HTS films in the small-signal limit and were very successful in optimising the quality of the films by careful adjustment of cation and oxygen stoichiometry [14]. Following a similar rationale, we have started a complementary research project that focuses on the nonlinear microwave response, including the

power- and temperature-dependent surface impedance and two-tone intermodulation distortion (IMD). We report here the first results of this project. We briefly introduce the techniques used for sample preparation and microwave measurements in Sec. II together with typical results. In all eight $YBa_2Cu_3O_{7-\delta}$ films on MgO that we have characterised so far, we have found a dramatic reduction of the surface resistance $R_s$ of up to one order of magnitude at microwave currents $I_{rf}$ in the mA-range for temperatures below 20 K. The surface reactance shows only a weak increase in the same current range. The anomalous decrease of $R_s(I_{rf})$ correlates with a pronounced plateau in the IMD signal. In Section III, we present a phenomenological model developed to describe our experimental data. It is based on a two-fluid model incorporating a microwave current-dependent quasiparticle scattering rate $g$ and normal fraction $f_n$ and involves only three adjustable parameters. We calculate the surface impedance and the third-order intermodulation product and find good agreement with the measured data. Section IV contains a detailed comparison between data and model and discusses the microscopic mechanisms that could cause the observed anomalous behaviour. The major results are summarised in Sec. V.

## II. EXPERIMENTAL DETAILS

A. Sample preparation and characterisation

Epitaxial $YBa_2Cu_3O_{7-\delta}$ (YBCO) films were prepared by electron-beam coevaporation onto homoepitaxially buffered 1 cm × 1 cm MgO substrates [15]. The 350-nm-thick single-sided films were deposited within a few percent of the 1:2:3 cation stoichiometry as judged from their morphology and energy dispersive X-ray analysis [14,16]. Oxygen content and ordering were adjusted by repeated plasma annealing in argon or activated oxygen. The final oxygenation level was judged from the critical temperature $T_c$, critical current density $J_c$ and flux creep rate derived from DC magnetisation measurements, and from the c-axis lattice parameter obtained from X-ray diffraction. So far, we have investigated eight different films prepared in three batches with oxygen contents ranging from underdoped to overdoped. Typical parameters are $T_c$-values between 69 and 92 K and $J_c$-values between 1 and 4 MA/cm$^2$ at 77 K and between 2.7 and 11 MA/cm$^2$ at 60 K, indicating a high sample quality. The XRD data yielded rocking-curve widths $\Delta W$ of the YBCO (005) reflection between 0.4 and 0.7 degrees. The c-axis parameters increased with oxygen deficiency $\delta$ as expected, with typical values ranging between 1.168 and 1.175 nm.

The microwave surface impedance and third-order two-tone intermodulation distortion were measured using the stripline-resonator technique developed at MIT Lincoln Laboratory [17]. The films were patterned by wet etching in dilute phosphoric acid into 150-µm-wide meander lines, and flip-chip packaged with two separate YBCO ground planes on MgO to form the stripline structure. The characteristic impedance of the resonator was 47 Ω and the frequency of the fundamental mode was $\omega/2\pi = 2.27$ GHz, where $\omega$ is the angular frequency. Computer-controlled measurements of the loaded quality factor $Q$ and the resonant frequency $\omega$ were performed in the fundamental resonance and four overtones for input power levels between −85 and +30 dBm, corresponding to a variation of the total RF current $I_{rf}$ in the stripline between 5 µA and 5 A. The surface resistance $R_s$ and changes of the surface reactance $X_s$ were derived from $Q$ and $\omega$ using the numerical procedure described in [18]. Computer-controlled IMD measurements with a noise floor of −130 dBm were performed for a frequency separation of 10 kHz of the two input tones. The input power of the two signals was kept equal, resulting in a symmetric IMD signal centered around the resonant frequency. The measurements were performed either in a flow cryostat or in a sealed package immersed into a liquid-helium dewar. The temperature could be adjusted between 1.7 K and 4.2 K by varying the vapour pressure of the helium and between 4.2 K and $T_c$ using a resistive heater.

We present in this report the detailed results for one representative film, which was slightly underdoped ($T_c = 90.7$ K, $J_c$(77K) = 1.5 MA/cm$^2$, $J_c$(60K) = 5.8 MA/cm$^2$, $c = 1.1725$ nm, and $\Delta W$=0.527 degrees). Typical $R_s$-values at $\omega/2\pi = 2.3$ GHz were 60 µΩ at $T = T_c/2$ and 2 µΩ at $T = 5$ K. At temperatures $T > 20$ K the surface resistance remained constant up to $I_{rf} = 0.5$ A, which corresponds to a magnetic flux density of about 15 mT. This result confirms the high structural and electronic quality of the film and suggests that weak Josephson coupling at grain boundaries is unlikely. The microwave measurements yielded qualitatively similar results for all other films, with a trend of weaker anomalies in more heavily doped films. The quantitative differences related to the different oxygenation levels will be described separately in a later publication.

B. Typical results for the surface impedance at low temperatures, $T < 20$ K

Figure 1 displays the temperature dependence of the surface resistance of our reference sample measured in the fundamental mode for the two levels of the input power that produce extreme $R_s$-values, $-60$ dBm (hatched squares) and $-20$ dBm (dots). The total RF current corresponding to these power levels is about 5 µA and 100 mA, respectively. Note that the temperature and resistance scales are logarithmic for illustration. Above $T = 20$ K, the two curves are indistinguishable. We define $T_{ano}$ as the temperature below which the anomalous effects occur, $T_{ano} = 20$ K here. At lower temperatures, $R_s$ increases with cooling at low power, but at the higher power level drops abruptly and approaches a constant low value of 2 µΩ. At still higher input power, as shown for $T = 5$ K by the dots in Fig. 2, $R_s$ starts to increase again due to the usual nonlinear behaviour, which may be related to penetration of flux into the stripline [19-21].

Figure 2 shows the microwave current dependence of the surface resistance and change of surface reactance at the constant temperature $T = 5$ K and $\omega/2\pi = 2.3$ GHz. $R_s(I_{rf})$ passes through a minimum at intermediate current levels, $I_{rf} \sim 100$ mA, where $R_s$ is reduced by approximately a factor of five compared with the low-current value. In contrast to the surface resistance, the reactance (crosses in Fig. 2) remains independent of power in the region of anomalous $R_s(I_{rf})$. This important result reveals that pair breaking or other sources leading to a current-dependent normal fraction $f_n(I_{rf})$ cannot explain our data. In the context of the two-fluid model, any explanation must therefore involve a current-dependent variation of the quasiparticle scattering rate, as discussed in detail in Sec. III.

The anomalous nonlinear behaviour is further illustrated in Fig. 3 in terms of $R_s(I_{rf})$ for various temperatures ($T$ = 1.7 to 20 K, panel *a*) and frequencies ($\omega/2\pi$ = 2.3 to 11.2 GHz, panel *b*). The anomalous current-induced reduction of the microwave losses becomes more pronounced at lower temperatures and disappears at high frequencies. There is no indication of saturation at the lowest temperature, $T$ = 1.7 K, in accordance with the finite slope of $R_s(T)$ at low power (Fig. 1). From the frequency-dependent data, we conclude that the surface resistance scales like $R_s(\omega,I_{rf}) \propto \omega^{\kappa(I)}$ with an effective current-dependent exponent $\kappa(I_{rf})$

$$\kappa(I_{rf}) = \frac{\partial \log R_s(I_{rf})}{\partial \log \omega}. \tag{1}$$

The data in Fig. 3b reveal $\kappa(0) \approx$ 1.4 to 1.5 at $T$=5 K, and $\kappa$ = 1.9 to 2 in the region of minimal $R_s$, independent of temperature in the entire range $T \leq T_{ano}$. This means that the minimum level of $R_s$ scales with frequency as expected from the usual two-fluid model (see, e.g., [3] and Sec. III.), while $R_s$ at the lowest currents displays a weaker frequency dependence.

Figure 4 compares typical results for the current-dependent $R_s$ and $X_s$ with the third-order IMD product. To avoid subtleties related to the conversion of input power into RF current for the two-tone experiment, the ordinate displays the nominal input power per tone. The data show the usual behaviour of the surface reactance, which increases smoothly with increasing power. The IMD signal displays at low power an approximately cubic power-law behaviour as expected for adiabatic non-hysteretic nonlinearities [3,22,23]. It passes through a pronounced plateau in the power range where $R_s$ decreases, and becomes again approximately cubic at higher power where both $X_s$ and $R_s$ increase. Figure 4 hence indicates that the anomalous behaviour of the IMD signal is related to that of the nonlinear surface resistance rather than the reactance. This is an unexpected result because the contribution of the nonlinear reactance to the IMD product is usually enhanced by the large ratio $X_s/R_s \gg 1$ [22].

Figures 1 to 4 represent typical results for all samples investigated to date. In addition to our own results, there are numerous reports of anomalous nonlinear microwave response in YBCO films in the literature (see [3,7-13] and references therein), indicating the necessity to understand the origin of these phenomena.

### III. TWO-FLUID MODEL WITH CURRENT-DEPENDENT SCATTERING RATE

A. Application of the two-fluid model to the nonlinear surface impedance

The two-fluid model (TFM) is now well established to describe the linear microwave response of high-temperature superconductors (e.g., [24-26]). Various groups have successfully applied this model to extract the temperature-dependent quasiparticle fraction $f_n$ and reduced scattering rate $g = (\omega\tau)^{-1}$ from surface impedance data, where $\tau^{-1}$ is the quasiparticle scattering rate. One especially important result was the observation of a linear variation of the normal fraction at low temperatures, which revealed the predominant d-wave character of the layered cuprate superconductors [27,28].

Here we extend the TFM to the anomalous nonlinear microwave response of YBCO. While this approach is purely phenomenological, it proves to be a valuable technique to describe our experimental data, and to derive important conclusions about the possible underlying microscopic mechanisms. We assume that both TFM variables, $g$ and $f_n$, depend in general on microwave current. From our experimental data in Fig. 2 we expect that $g(I_{rf})$ will be important for the anomalous behaviour, while $f_n(I_{rf})$ can be relevant only for the nonlinear surface impedance at high currents. The most prominent sources of a current-dependent normal fraction are pair breaking [22] and granularity [29,30]. Reasons for a current dependent quasiparticle scattering rate are less obvious. Possible sources are strong electronic correlations, which could lead to a simultaneous increase of $f_n$ and $g$, e.g., in the case of pair breaking. Other mechanisms are discussed in Sec. IV.E. Table I summarises the assumptions and parameters of our model.

It is reasonable to assume that the quasiparticle scattering rate depends only on the magnitude of the microwave current or magnetic field, $h$, but not on its direction. We assume for simplicity a quadratic field dependence of $g$:

$$g = g_0 \times \left[1 + h \times \frac{h^2}{(1+h^2)}\right], \qquad (2)$$

where $h$ is the microwave current or field, normalised to an appropriate scale. The anomalous current dependence of $R_s$ described in Sec. II.B. reveals typical scaling currents in the mA-region, corresponding to RF magnetic flux densities around 30 µT. Equation (2) is the simplest construction of a field dependent scattering rate, which involves two parameters: $g_0$ the residual scattering rate at $h = 0$ and $h$ the total relative variation of $g$, $h = g_\infty/g_0 - 1$, where $g_\infty = g(h \to \infty)$.

The second variable in the two-fluid model is the normal fraction $f_n$. In a similar way, we assume that $f_n$ increases quadratically with magnetic field, which is in accordance with the phenomenological model developed in [22]:

$$f_n = f_{n0} + (1 - f_{n0}) \times \frac{(\beta h)^2}{1 + (\beta h)^2}, \qquad (3)$$

with $f_{n0}$ the residual fraction of unpaired charge carriers at $h = 0$. The prefactor and the field dependence of the second term assure that $f_n$ saturates at the normal-state value, $f_n(h \to \infty) = 1$. The scaling factor $b$ has to be introduced to distinguish the pair-breaking effect from the field-dependent quasiparticle scattering rate. Larger values of $b$ lead to earlier onset of pair breaking. The experimental data show a nonlinear increase of the surface impedance in the range of 500 mA (or 15 mT), corresponding to $b \sim 0.002$.

B. Surface impedance

The surface impedance $Z_s = R_s + i X_s$ in the framework of the TFM is completely determined by $f_n$ and $g$ since the superconducting pair fraction is related to $f_n$ by $f_s = 1 - f_n$. $Z_s$ can be calculated rigorously in the local limit, which applies to HTS [31], $Z_s = (i\mu_0 \omega/\sigma)^{1/2}$, with $\sigma$ the complex Drude-type conductivity $\sigma = \sigma_1 - i\sigma_2$ [28]:

$$\mu_0 \omega \lambda_0^2 \times \sigma = f_n \frac{g^{-1}}{1+g^{-2}} - i\left[ f_n \frac{g^{-2}}{1+g^{-2}} + (1-f_n) \right] \tag{4}$$

Solving the complex root for $Z_s$ yields for arbitrary values of $f_n$ and $g$ (e.g. [3,28]):

$$z \equiv \frac{Z_s}{\mu_0 \omega \lambda_0} = i \times \frac{(1+g^{-2})^{1/2}}{[f_n^2 g^{-2} + (1-f_n + g^{-2})^2]^{1/4}} \times \exp[-\frac{i}{2} \tan^{-1}(\frac{f_n g^{-1}}{1-f_n + g^{-2}})]. \tag{5}$$

The parameter $\lambda_0$ denotes the minimum possible penetration depth for complete pairing, $f_s = 1$. The value of $\lambda_0$ is difficult to derive from experiment [32,33]. However, it is used in Eq. (5) only for normalisation, i.e., absolute $\lambda_0$-values are irrelevant for our purpose. The field dependence of $Z_s$ is implicitly contained in $g(h)$ and $f_n(h)$, Eqs. (2) and (3). It is instructive to expand Eq. (5) for small values of $f_n$ and $g^{-1}$:

$$r \equiv \frac{R_s}{\mu_0 \omega \lambda_0} \approx \frac{1}{2} f_n g^{-1} \times (1 + \frac{3}{2} f_n) \quad \text{and} \quad x \equiv \frac{X_s}{\mu_0 \omega \lambda_0} \approx 1 + \frac{1}{2} f_n + O(g^{-2}) \tag{6}$$

Equation (6) reproduces the quadratic and linear frequency dependences of $R_s$ and $X_s$ and the linear scaling of $Z_s$ with the normal fraction $f_n$, which are typical for the usual two-fluid model. Another important result is the linear dependence of $R_s$ on the scattering time $\omega \tau = g^{-1}$, whereas $X_s$ is independent of $g$ in this limit.

C. Derivation of $f_n$ and $g$ from the surface impedance

Equation (5) implies that $f_n$ and $\gamma$ are known quantities, which is in contrast with the experimental situation where $Z_s$ is obtained from measured data. The TFM variables can be related to the surface impedance via the complex conductivity, Eq. (4), by:

$$\frac{1}{\mu_0 \omega \lambda_0^2} \times (1 - f_n) = \sigma_2 - \gamma^{-1}\sigma_1. \tag{7}$$

In turn, the complex conductivity can be derived from the surface impedance by

$$\sigma = \frac{\mu_0 \omega}{\left(X_s^2 + R_s^2\right)^2} \times \left[2 R_s X_s - i\left(X_s^2 - R_s^2\right)\right]. \tag{8}$$

We can solve for $f_n$ and $\gamma^{-1}$ by combining Eqs. (7) and (8):

$$1 - f_n = \frac{X_{s0}^2 \left(X_s^2 - R_s^2 - 2\gamma^{-1} R_s X_s\right)}{\left(X_s^2 + R_s^2\right)^2} \quad \text{and} \quad \frac{\gamma^{-1}}{1 + \gamma^{-2}} = \frac{2 R_s X_s X_{s0}^2}{\left(X_s^2 + R_s^2\right)^2} \times \frac{1}{f_n}, \tag{9}$$

where $X_{s0} = \mu_0 \omega \lambda_0$. The second part of Eq. (9) is quadratic in $\gamma^{-1}$ with the positive root

$$\gamma^{-1} = \frac{2 R_s X_s X_{s0}^2}{\left(X_s^2 + R_s^2\right)^2 - X_{s0}^2\left(X_s^2 - R_s^2\right)}. \tag{10}$$

The results in Eqs. (9) and (10) can be simplified for the usual limit $X_s \gg R_s$, leading to the approximate expressions in reduced variables:

$$f_n(h) \approx 1 - \frac{1}{x^2(h)} \quad \text{and} \quad \gamma^{-1}(h) \approx \frac{2 r(h)}{x(h)} \times \frac{1}{x^2(h) - 1}, \tag{11}$$

where the field dependences are explicitly stated. Accurate determination of absolute values of $f_n$ and $\gamma$ requires knowledge of absolute values of $x$ and hence of $\lambda$ and $\lambda_0$, which are, in principle, difficult to determine. However, in cases like ours, where $x$ is almost independent of microwave current, the relative variation of the scattering rate is uniquely determined by that of the surface resistance (Sec. IV.B.).

D. Third-order two-tone intermodulation product

We start with a *qualitative* study of the IMD signal expected for the two-fluid model with a current-dependent scattering rate, based on an expansion of $Z_s$ in terms of $\gamma^{-1}$ and $f_n$, Eq. (6). The reduced surface impedance then becomes a rational function containing only even powers of $h$:

$$r \approx \frac{A + B \times h^2 + C \times h^4}{1 + (\boldsymbol{h} + 1) \times h^2} \tag{12}$$

and $x \approx x_0$ = constant. The denominator is reminiscent of the current-dependent scattering rate, Eq. (2). The coefficients $A$, $B$ and $C$ depend on $f_{n0}$, $\boldsymbol{g}_0^{-1}$, and $\boldsymbol{b}$ but are independent of $\boldsymbol{h}$. Their values determine obviously the existence, position and depth of a minimum of the nonlinear surface resistance $r(h)$.

To estimate the third-order IMD signal, we note that the electric field $e$, in dimensionless units, is related to the magnetic field by $e = (r + ix) \cdot h$. We assume a power series for $e$, $e = \sum_{n=0}^{\infty} \boldsymbol{e}_n h^n$, and find the coefficients $\boldsymbol{e}_n$ from the resulting equation $\boldsymbol{e}_0 + \boldsymbol{e}_1 h + (\boldsymbol{h} + 1) \sum_{n=0}^{\infty} (\boldsymbol{e}_n + \boldsymbol{e}_{n+2}) h^{n+2}$ $= (A + ix_0)h + (B + ix_0)h^3 + Ch^5$. For a symmetric two-tone signal, $h = h_0 \cdot (\cos\boldsymbol{w}_1 t + \cos\boldsymbol{w}_2 t)$, it is possible to derive an analytical expression for the scaling of the electric field component with the field amplitude $h_0$ at the two side-bands, $f_{+/-} = 2f_{1/2} - f_{2/1}$, which complements the formal result given in [3]:

$$e_{IMD} \propto h_0^3 \frac{1}{1 + 4h_0^2} . \tag{13}$$

The factor 4 in the denominator is generic of the *third* order of the intermodulation. The corresponding magnetic field amplitude induced by the intermodulation is given by

$$h_{IMD} = \frac{|e_{IMD}(h_0)|}{|z(h_0)|} . \tag{14}$$

Given that the denominator of Eq. (14) is dominated by the constant reactance, $|z(h_0)| \approx x_0$, we expect a cubic and linear asymptotic behaviour, $h_{IMD} \propto h_0^3$ and $h_{IMD} \propto h_0$, at low and elevated power, respectively.

While the above treatment was based on series representations of the surface impedance and the electric field, numerical algorithms are required to calculate $h_{IMD}$ in the intermediate power range or for arbitrary values of field level and model parameters. We have developed such a computer algorithm whose results are discussed in the following Section.

## IV. RESULTS AND DISCUSSION

A. Anomalous nonlinear surface impedance and IMD product

Figure 5 displays typical results for the *nonlinear* two-fluid model for $R_s$, $X_s$, and IMD signal for typical parameter values ($g_0^{-1} = 0.01$, $f_{n0} = 0.1$, $h = 5$) and different scaling parameters $b$ ($b = 0$, 0.01, 0.1, 1.0 for the solid, long-dashed, dashed, and dotted curves). The anomalous nonlinear behaviour changes only quantitatively but not qualitatively with the values of $g_0$, $f_{n0}$, or $h$ set within realistic ranges. It is worth noting that $g_0^{-1}$ and $f_{n0}$ cannot be chosen arbitrarily but are fixed by the asymptotic value of the surface impedance at the lowest currents.

Figure 5a shows the surface resistance computed from Eq. (5). The $R_s$ starts to increase at progressively lower field levels for increasing $b$-values. The field range and the depth of the anomalous behaviour shrink and vanish for $b \geq 1$, where the pair breaking begins dominating the current-dependent quasiparticle scattering. The absence of an anomalous field dependence in the reactance (Fig. 5b) is a characteristic feature of our model by design and reflects that $X_s$ is independent of the quasiparticle scattering rate [see Eq. (6)]. Like the surface resistance, $X_s$ increases at progressively lower field levels as $b$ increases above zero. The maximum of $X_s$, which occurs for $b \geq 1$, is a generic feature of the two-fluid model. The saturation of $X_s$ at the highest field levels indicates complete depairing, $f_n = 1$. It can be seen that the parameter $b$ is merely needed to adjust the high-field nonlinearities but does not affect the current-dependent scattering rate, which causes the anomalous behaviour.

Figure 5c shows the IMD product, which was numerically computed on the basis of the full expressions for the electric field and the surface impedance, Eqs. (5) and (14). We find the asymptotic behaviour expected from Eq. (13) confirmed for $b = 0$. Moreover, the calculated IMD signal displays a plateau, which resembles the experimental results described in Sec. II.B., except for the high-power side where the measured IMD signal increases more strongly than linear (squares in Fig. 4). This difference can be attributed to the field-scaling parameter ***b***: For intermediate ***b***-values, the IMD signal displays a transition from linear to cubic high-field behaviour, as shown by a comparison between the solid and long-dashed curves in Fig. 5c. At high power and for ***b*** ≥ 1, another linear region develops, which results from the nonlinear increase of the surface impedance, $z \propto h_0^2$.

Figure 6 characterises the plateau of the calculated IMD product between the two cubic regions for moderate ***b***-values in greater detail. The width of the plateau, $h_2 - h_1$, normalised to the lower edge $h_1 = 0.1$, decreases with increasing field-scaling parameter ***b*** approximately like ***b***$^{-0.85}$ up to ***b*** ~ 0.01 and vanishes towards ***b*** ~ 0.1 as indicated by the arrow. The width of the measured IMD plateau can hence be used to fix the parameter ***b***. According to figure 4, a typical IMD plateau extends over about 30 dB in power, which corresponds to a reduced width of √1000 ~ 30 and thus a field-scaling parameter ***b*** ~ 0.003. This value is consistent with the experimental results for the surface impedance described in Sec. II. It justifies our assumption of different field scales for pair breaking (increasing $f_n$) and scattering (increasing ***g***). It underlines at the same time the different sources of the $R_s$ minimum and IMD plateau on one hand and the high-power nonlinearities on the other hand.

B. Current-dependent quasiparticle scattering rate

We derive the microwave current-dependent scattering rate expected for the *nonlinear* two-fluid model from the measured $Z_s$ data as described in Sec. III.C. We start with the data for $T$ = 5 K and ***w/2p*** = 2.27 GHz. We use the simplified expressions for $f_n$ and ***g***, since $R_s$ ~ μΩ << $X_s$ ~ mΩ [see Eq. (11)]. According to our previous discussion, the absolute values of $f_n$ and ***g*** depend sensitively on the values of ***l***($I_{rf}$=0) and ***l***$_0$. However, our experimental data reveal that $X_s(I_{rf})$ is constant with %-accuracy in the current range of the anomalous $R_s(I_{rf})$. We focus on this range because the nonlinearities at higher currents seem to be caused by a different mechanism, possibly flux penetration into the stripline [20,21]. With $x(h)$ = constant, we can accordingly normalise Eq. (11) so that $f_n(I_{rf}) / f_n(0) = 1$ and $g(I_{rf}) / g(0) = R_s(0) / R_s(I_{rf})$.

The dots in Fig. 7 represent the inverse normalised surface resistance, or the reduced scattering rate, derived from the data in Fig. 3. We can clearly identify asymptotic behaviour at low and high currents, giving us the opportunity to compare the measured results with the field dependence of the scattering rate $g$ assumed in Eq. (2) (solid curve in Fig. 7):

$$\frac{g(I)}{g(0)} = 1 + h\frac{(I/I_0)^2}{1+(I/I_0)^2}. \tag{15}$$

We had to introduce the scaling current $I_0$ in Eq. (15) for an explicit comparison between data and model [$I/I_0$ replaces the reduced variable $h$ in Eq. (2)]. The two model parameters are found to be $h = 1.65$ and $I_0 = 7$ mA. We note slight differences between data and model in the transition regions, which could be affected by the non-uniformity of the field across the stripline or by the mechanism that causes the increase of $R_s(I_{rf})$ at high currents. A more sophisticated choice of the current dependence might reduce this discrepancy, though at the expense of additional model parameters.

Similar agreement between model and experimental results was found at the higher frequencies, as illustrated in Fig. 7 by the different symbols and fits, and throughout the entire temperature range below $T_{ano}$. The fits each use different values of $h$ and $I_0$ since these parameters are frequency and temperature dependent. Keeping in mind the phenomenological nature of the parameters (see Sec. IV.D. for further discussion), we can apply our model to reduce the measured data and map them onto the temperature dependences of $h$ and $I_0$. Figure 8 summarises the results for the temperature range 1.7 to 20 K. With increasing temperature $h$ decreases and vanishes at $T = T_{ano} \approx$ 20 K, in accordance with the power splitting of $R_s(T)$ observed in Fig. 1. The temperature dependence of $h$ parametrises the thermal variation of the depth of the $R_s(I_{rf})$ minimum. Our observation of the surface resistance at low power continuing to increase at the lowest temperatures is reflected by the divergence of $h(T\rightarrow 0)$ or, more precisely, by a divergence of $g^{-1}$ at zero temperature.

The scaling current $I_0$ starts at the order of 1 mA and increases with temperature towards $T_{ano}$, where it reaches values between 100 and 1000 mA. This $I_0(T)$ increase reflects the shift of the $R_s$ minimum to higher currents and the weakening of its current dependence as $T$ increases. While the latter is accordance with the decreasing $h$ values, $h$ and $I_0$ are in general independent parameters. The nonlinear increase of $R_s$ and $X_s$ at high currents dominates the anomalous nonlinear response at elevated temperatures so that the minimum of $R_s$ disappears above $T = 20$ K.

C. Frequency dependent quasiparticle scattering rate

In the framework of our model, the variable frequency dependence of $R_s$ at low currents is strictly related to that of $(1+h)$, since the surface resistance is proportional to the scattering rate. The frequency exponent $k(I_{rf}\rightarrow 0)$ of Eq. (1) can hence be expressed by

$$k(0) = 2 + \frac{\partial \log g_0^{-1}}{\partial \log w} = 2 + \frac{\partial \log (1+h)}{\partial \log w}. \qquad (16)$$

The second equality uses the identity $g_0^{-1} = (1+h) \times g_\infty^{-1}$ and the frequency independence of $g_\infty$, which we concluded from the quadratic frequency dependence of $R_s(I_{rf})$ in the region of the minimum. Equation (16) reveals that $1+h$ decreases with increasing frequency, because $k(0)$ must remain smaller than two to fit the measured data. If we formally write $g_0^{-1} = g_\infty^{-1} + h \times g_\infty^{-1}$ and assume an effective power-law behaviour for the frequency dependence of $h$ itself, $h(w) \propto w^{b_h}$ with $b_\eta < 0$, we can avoid the logarithmic derivative in Eq. (16) and obtain instead

$$k(0) = 2 - |b_h| \times \frac{h}{1+h}. \qquad (17)$$

This identity indicates that the frequency dependence of the low-current surface resistance becomes weaker for stronger anomalies, i.e., for larger **h**-values. This direct consequence of the two-fluid model is in accordance with our data (Fig. 3) and confirms that the scattering rate in the anomalous regime is indeed frequency dependent.

Figure 7 illustrates the fits to the frequency-dependent $R_s(I_{rf})$ data at $T = 5$ K, which confirm the expected decrease of **h** with increasing frequency. The transition region between the asymptotic levels at low and high currents remains unchanged, i.e., $I_0(w)$ remains approximately constant. This behaviour could be confirmed for the entire temperature range, as illustrated for 1+**h** in figure 9. We noted a weak trend for the scaling current to decrease with increasing frequency at temperatures below $T = 5$ K. However, the $I_0$-variation remained still below a factor of 2 and is not considered significant.

The frequency dependence of 1+**h** becomes shallower as the temperature approaches $T_{ano}$. We have analysed the frequency exponent of (1+**h**), as indicated in Eq. (16) and illustrated in the inset to Fig. 9, and found that it is consistent with approaching the value $-1$ at $T \to 0$ K. This value corresponds to a linear frequency dependence of $R_s$, which, in turn, means that the scattering rate of the anomalous mechanism becomes proportional to frequency at zero temperature.

D. Microscopic mechanisms

The major findings of our studies are: A current-independent normal fraction $f_n$, a current-dependent scattering rate **g** whose magnitude increases almost linearly with temperature and frequency, and a scaling current that strongly increases with temperature. We now discuss briefly possible microscopic mechanisms that could cause the observed behaviour.

We start by considering Josephson coupling across grain boundaries, which was suspected to be responsible for anomalous nonlinear microwave response of high-temperature superconductors [3,9,34-36]. We can exclude this mechanism in our case for the following reasons. First, the scaling current $I_0(T)$ drops at low temperatures, which is in contrast to the temperature dependence of the critical Josephson current density $J_{cJ}$. Second, the measured nonquadratic frequency dependence of $R_s$ at low currents contradicts the expectation for Josephson junctions, which display $R_s \propto w^2$ for sufficiently high $I_c R_n$-values and $J < J_{cJ}$, and $R_s \propto w^{1/2}$ for $J \gg J_{cJ}$. Finally, our samples display excellent power handling at elevated temperatures, which is in contrast to the granular behaviour expected for weakly coupled grains [30].

Nonlocal effects were considered responsible for the weak temperature dependence of the surface resistance below $T = 40$ K [31]. This is in contrast to our observation of a strong temperature dependence of $R_s$ down to the lowest temperatures studied, which would reflect a strongly temperature dependent quasiparticle mean free path if this effect were due to nonlocality. The effective quasiparticle scattering rate would also not be expected to assume an explicit frequency dependence [37]. Furthermore, the product $g^{-1} f_n/(1-f_n)$ would need to be of the order of ~ 1, and hence $g^{-1} \sim 10$, to explain the weak frequency dependence of $R_s$ at low currents [3]. Such values are unrealistic for epitaxial films, and they also contradict the absence of power dependence of the surface reactance. Finally, nonlocal effects should be strongly suppressed in our measurement configuration [38,39].

Nonequilibrium effects can be another source of anomalous superconducting features including recovery effects [40]. Such phenomena, however, occur only above a critical frequency, which is of the order of the inelastic scattering rate [41]. This rate falls into the low THz region for the cuprate superconductors and hence is well beyond our measurement frequency. There are further discrepancies between the observed anomalies and non-equilibrium effects in conventional superconductors, associated with their frequency dependence, growth at low temperatures, and absence of effects in $f_n$. It is not obvious if and how these conclusions could be affected by a nodal order parameter such as in the cuprates, and it would definitely be interesting to gain deeper insight into non-equilibrium effects in d-wave superconductors. But since the quasiparticle population is larger and the excitation energies are lower due to the presence of nodes, we would intuitively expect non-equilibrium effects to be even less relevant.

While the above three sources of anomalous behaviour can be excluded with high probability, other mechanisms cannot yet be ruled out. Among those are the ordering of magnetic impurities [3,42], the opening of a superconducting subdominant s-wave-like energy gap [43,44], or other peculiar transport phenomena related to the separation of spin and charge in quasi one-dimensional systems [45,46], intrinsic disorder [47,48], the formation of stripes [49], or novel quasiparticle concepts [50].

## V. CONCLUSIONS

We have reproducibly observed a strongly frequency and temperature dependent microwave-current-induced decrease of the surface resistance $R_s(I_{rf})$ of high-quality epitaxial YBCO films on MgO at temperatures below 20 K. In contrast to $R_s$, the surface reactance was nearly constant within the same current range. The third-order two-tone intermodulation product showed an extended plateau surrounded by regions where the IMD signal rose almost with the third power of the input signal. The anomalous IMD data are clearly related to the current-induced reduction of $R_s$. Our observations of anomalous $R_s(I_{rf})$ behaviour are qualitatively similar to numerous reports of other groups but quantitatively stronger. In detail, our studies concern the *low*-temperature anomalies identified recently in Ref. 36. There are also reports of anomalies at significantly *higher* temperatures (e.g., [3,9,36] and references therein) which might have a different origin, such as weak grain-boundary coupling.

To model our experimental data we have developed an extended two-fluid model incorporating a current-dependent quasiparticle scattering rate $g$ and normal fraction $f_n$. This model allows us to condense the great number of experimental details into three adjustable parameters, the scattering rate span $h$, the scaling current $I_0$, and the field-scaling parameter $b$. The resulting phenomenological picture can be summarised as follows:

1. At low temperatures, $T < T_{ano} \sim 20$ K, $h(T,w)$ decreases strongly with increasing temperature and frequency. This variation can be associated with a scattering rate which increases almost linearly with temperature and frequency. The corresponding frequency dependence of the surface resistance becomes almost linear at the lowest currents but remains quadratic for currents at and beyond the region of the $R_s$ minimum. The nonmonotonic current and temperature dependences of $R_s$ are related through the same mechanism causing the temperature dependences of $h$ and $I_0$.

2. $I_0(T)$ shows a trend opposite to that of $h(T)$: it decreases from a 100 mA-level at $T = 20$ K with decreasing temperature and saturates on the 1 mA-level at low temperatures. The scaling current is almost independent of frequency, especially for $T$ near $T_{ano}$.

3. The field-scaling parameter $b(T)$ is of the order of $10^{-3}$ at $T \leq 5$ K and increases with temperature at the same rate as $I_0(T)$, approaching $b(T_{ano}) \sim 1$. In addition to the weakening of the anomalous response at elevated temperatures, the high-power nonlinearities start dominating the current dependent surface impedance and intermodulation distortion. The smallness of $b$ at low temperatures reveals different mechanisms to be responsible for the nonlinear response at low and high currents. The former is attributed to an increase of $g$, the latter to an increase of $f_n$.

Obvious mechanisms leading to nonlinear microwave response like Josephson coupling across grain boundaries, nonlocal or nonequilibrium effects could be ruled out as sources of the anomalous behaviour. The rationale of our model is consistent with an equivalent parallel circuit for charge transport in the YBCO films. The anomalies could therefore reflect a two-band behaviour or even different kinds of charge carriers. Our continuing studies of the influence of oxygen order and stoichiometry will help to complement our present understanding of charge and spin ordering and transport phenomena in the low-dimensional cuprate superconductors.

**ACKNOWLEDGMENT**

We gratefully acknowledge valuable contributions from S. Anlage, T. Dahm, J. Derov, C. Gough, J. Halbritter, K. Irgmaier, N. Klein, P. Lahl, M. Lancaster, K. Numssen, Sang-Hoon Park, A. Porch, and R. Wördenweber. We also thank R. Konieczka and D. Baker for stripline resonator fabrication and G. Fitch for programming. This work has been funded in part by the Land Nordrhein-Westfalen (Germany), EPSRC and MOD (UK), and AFOSR (U.S.A.). Part of this material is based upon work supported by the European Office of Aerospace Research and Development, Air Force Office of Scientific Research, Air Force Research Laboratory, under Contract No. F61775-01-WE033.

Table I: Summary of the two-fluid model with a current-dependent quasiparticle scattering rate. The only adjustable parameters are $h$, $I_0$, and $b$.

| Parameter | Definition | Explanation |
| --- | --- | --- |
| Quasiparticle (qp) scattering rate | $t^{-1}$, Eq. (4) | Ingredient for a Drude-like conductivity. |
| Reduced qp scattering rate | $g=(\omega t)^{-1}$, Eq. (2) | Increases quadratically with microwave current from the residual value $g_0$ at $I/I_0 \ll 1$ to the asymptotic level $g_\infty$ at $I/I_0 \gg 1$. |
| Residual scattering rate | $g_0$ | Determined by $R_s(I=0)$ and $X_s(I=0)$. |
| Scattering rate span | $h = (g_\infty - g_0)/g_0$ | Sets the scale of the anomalous drop of $R_s(I)$. |
| Scaling current | $I_0$ | Sets the absolute current scale for the anomalous behaviour. |
| Normal fraction | $f_n$, Eq. (3) | Increases as $(bI)^2$ from the residual value $f_{n0}$ at $bI \ll 1$ to $f_n=1$ at $bI \gg 1$. |
| Residual normal fraction | $f_{n0}$ | Determined by $R_s(I=0)$ and $X_s(I=0)$. |
| Scaling parameter | $b$ | Adjusts the relative current scales of $g$ and $f_n$, reflecting the importance of pair breaking relative to current-dependent quasiparticle scattering. |
| Surface impedance | $Z_s$, Eqs. (5), (6) | Assumes local electrodynamics. |

**FIGURE CAPTIONS**

Figure 1. Temperature dependent surface resistance at $\omega/2\pi = 2.3$ GHz for two input power levels (−60 dBm: hatched squares, −20 dBm: dots).

Figure 2. Microwave-current dependent surface resistance $R_s$ (dots) and changes of the reactance $\Delta X_s$ (crosses) at $T = 5$ K and $\omega/2\pi = 2.3$ GHz.

Figure 3. Measured microwave current dependent surface resistance (*a*) at $\omega/2\pi = 2.3$ GHz for various temperatures: $T$ (K) = 1.7 (dots), 2.0 (hatched squares), 3 (diamonds), 5 (triangles), 10 (inverted triangles), 15 (shaded squares), 20 (crosses) and (*b*) at $T = 5$ K for various frequencies: $\omega/2\pi$ (GHz) = 2.3 (dots), 4.5 (triangles), 6.7 (diamonds), 9.0 (squares), 11.2 (inverted triangles).

Figure 4. Measured dependence of the third-order intermodulation (left ordinate, squares), surface resistance (right ordinate, dots) and change of surface reactance (right ordinate, triangles) at $\omega/2\pi = 2.3$ GHz and $T = 5$ K.

Figure 5. Simulated microwave field dependence of (*a*) the reduced surface resistance, (*b*) surface reactance and (*c*) third-order intermodulation signal for various values of $\beta$ (solid curve: 0, coarse-dashed: 0.01, dashed: 0.1, dotted: 1.0).

Figure 6. Dependence of the plateau width of the simulated third-order intermodulation signal on the parameter $\beta$ (further explanations are given in the text).

Figure 7. Measured and fitted microwave current dependence (symbols / curves) of the reduced scattering rate at $T = 5$ K for various frequencies: $\omega/2\pi$ (GHz) = 2.3 (dots / solid curve), 4.5 (triangles / long-dashed), 9.0 (squares / dashed), 11.2 (inverted triangles / dotted). The fits yield the two parameters $\eta$ and $I_0$.

Figure 8. Temperature dependence of the model parameters $\eta$ (left ordinate, hatched squares, solid interpolation) and $I_0$ (right ordinate, dots, dotted interpolation) for $\omega/2\pi = 2.3$ GHz.

Figure 9. Frequency dependence of the model parameter $\eta$ for various temperatures: $T$ (K) = 1.7 (dots), 2.0 (hatched squares), 3 (diamonds), 5 (triangles), 10 (inverted triangles), 15 (shaded squares), 20 (crosses). The curves represent power-law fits to the data. The inset shows the temperature dependence of the slopes, $\kappa(0)-2$ [see Eq. (15)].

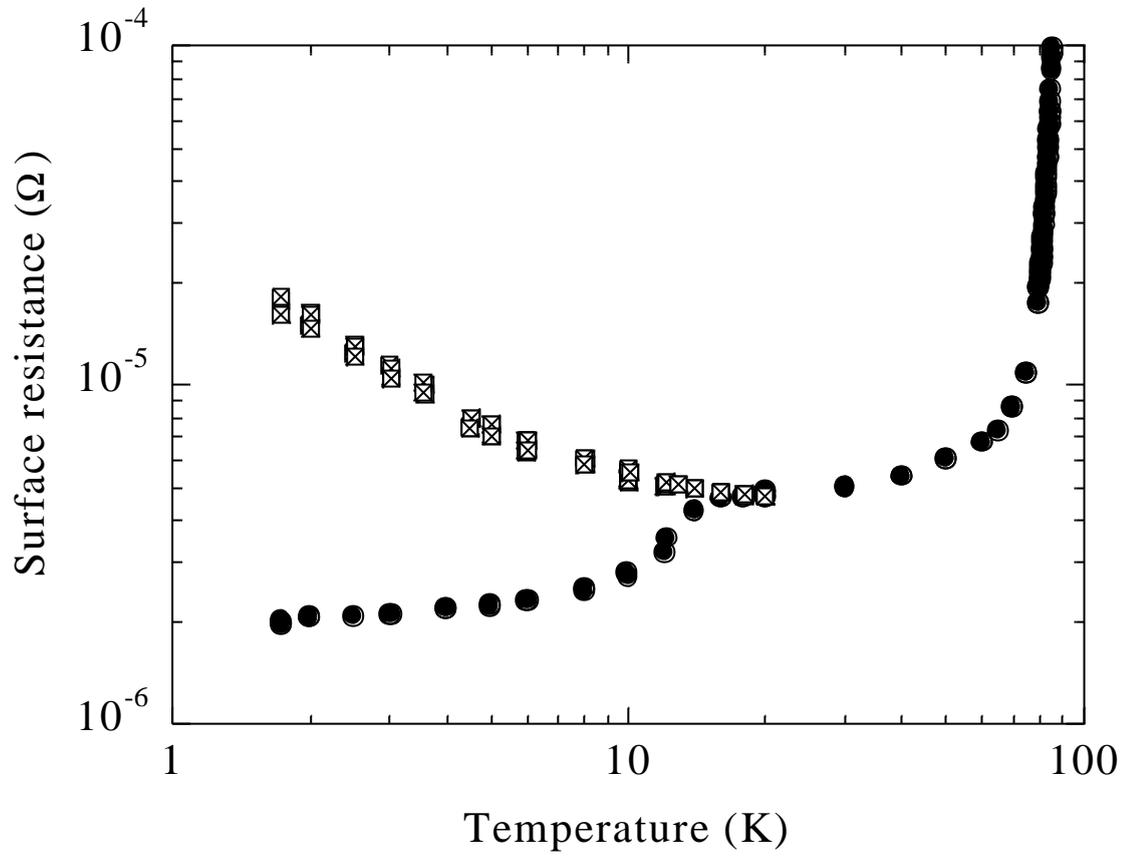

Figure 1.

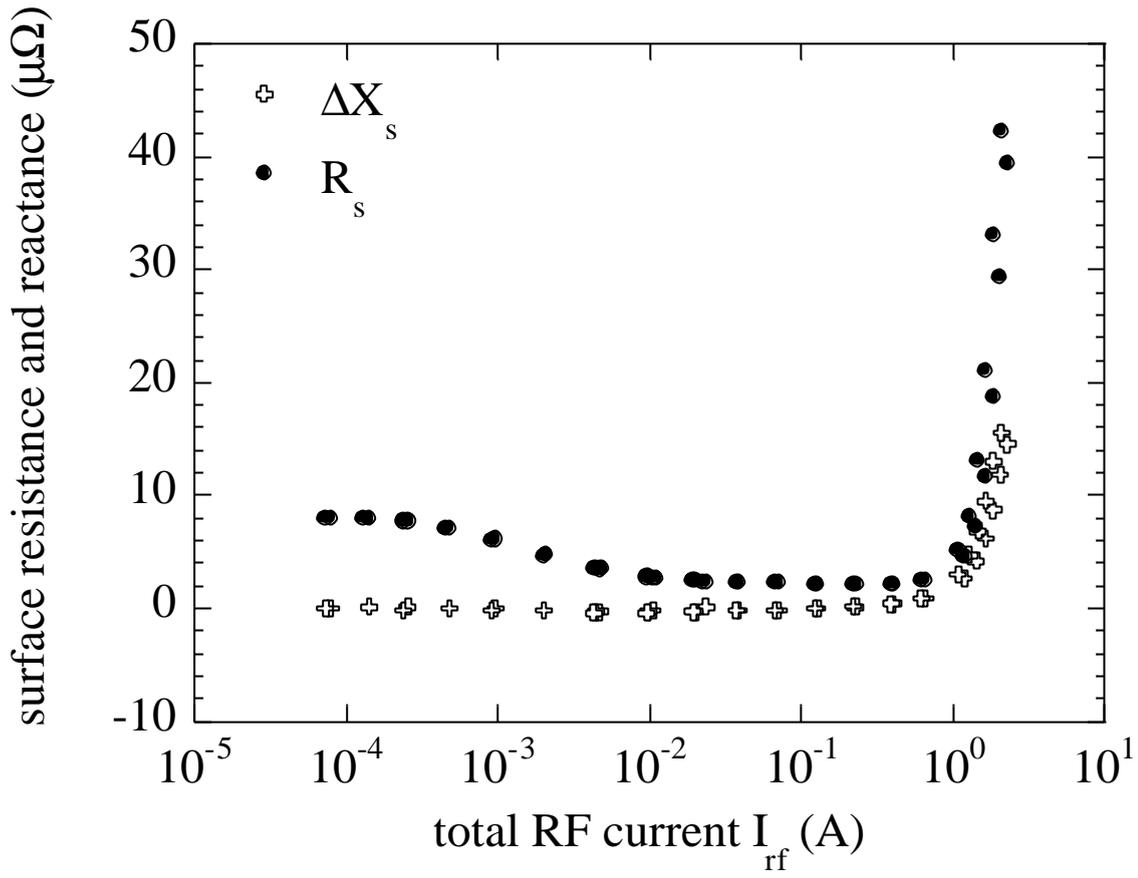

Figure 2.

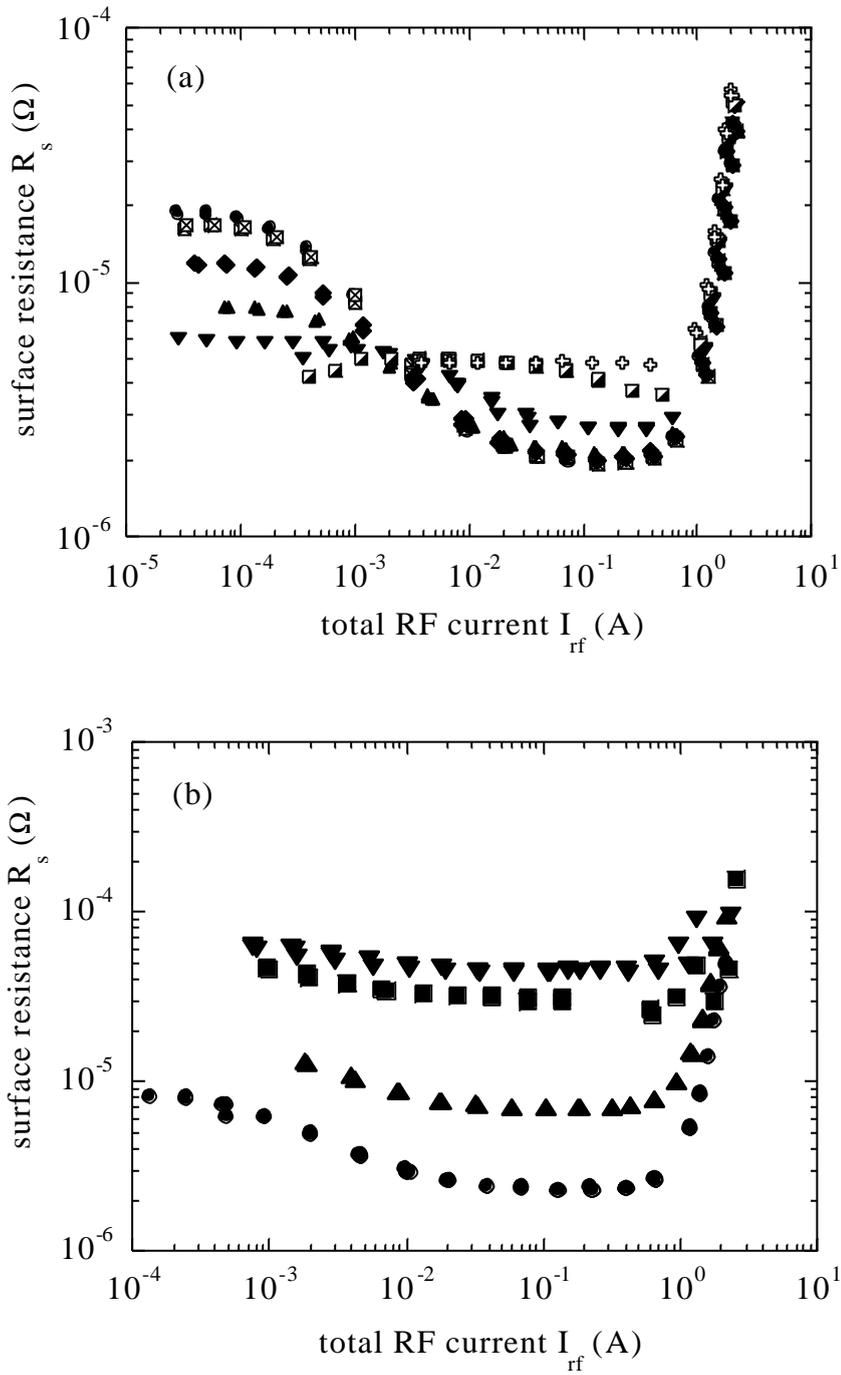

Figure 3.

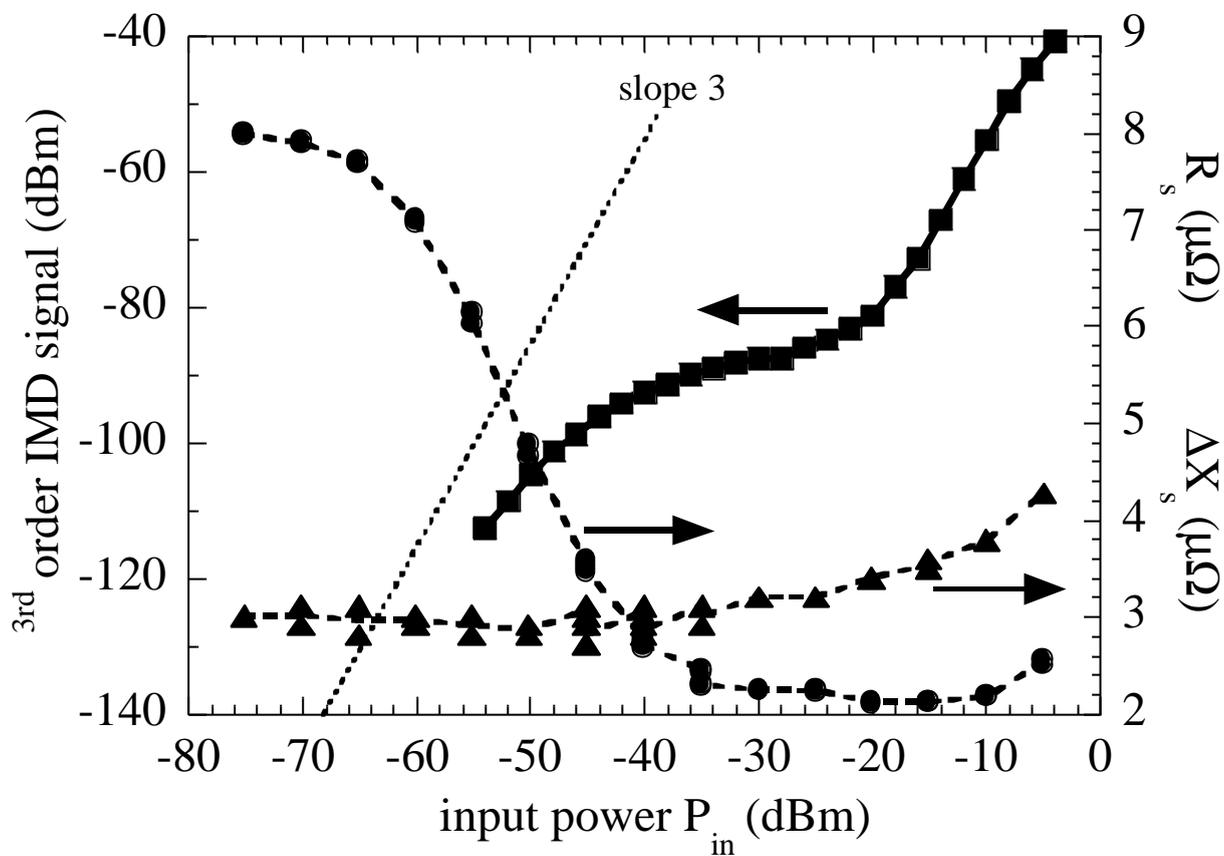

Figure 4.

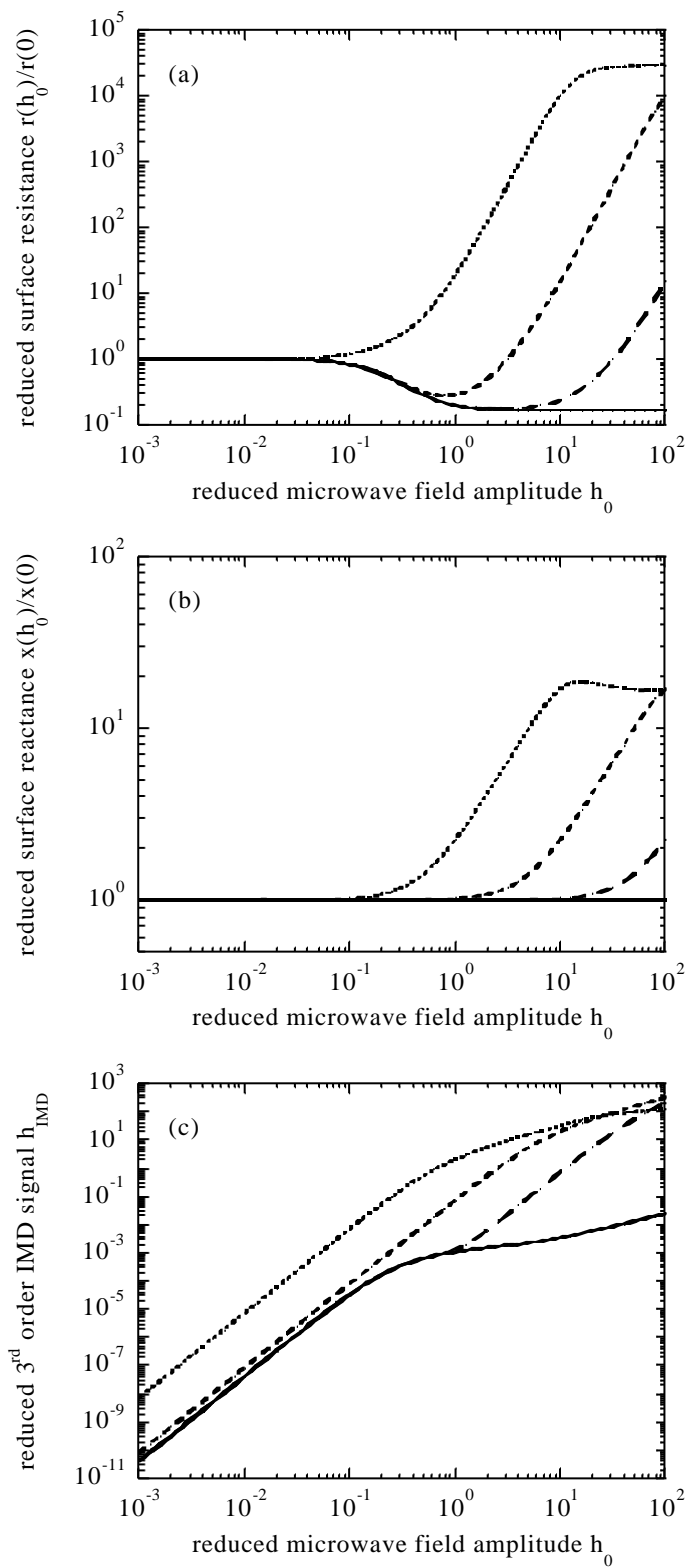

Figure 5.

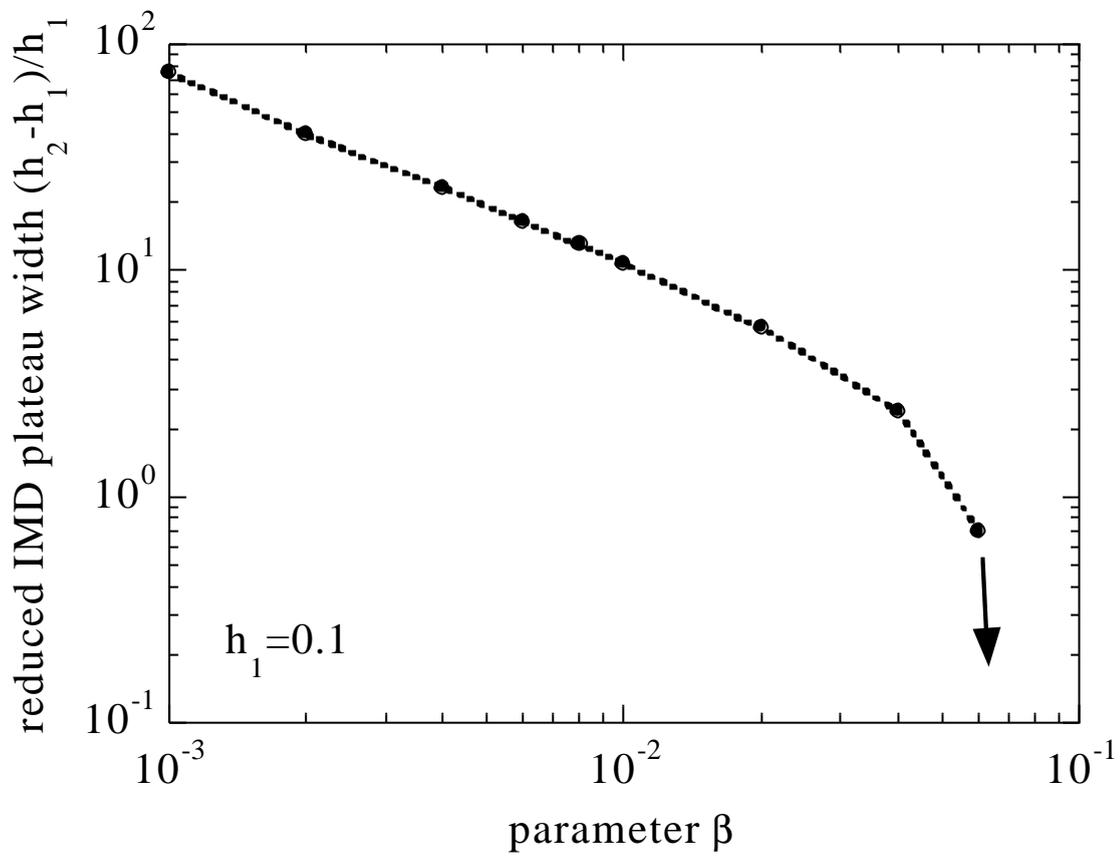

Figure 6.

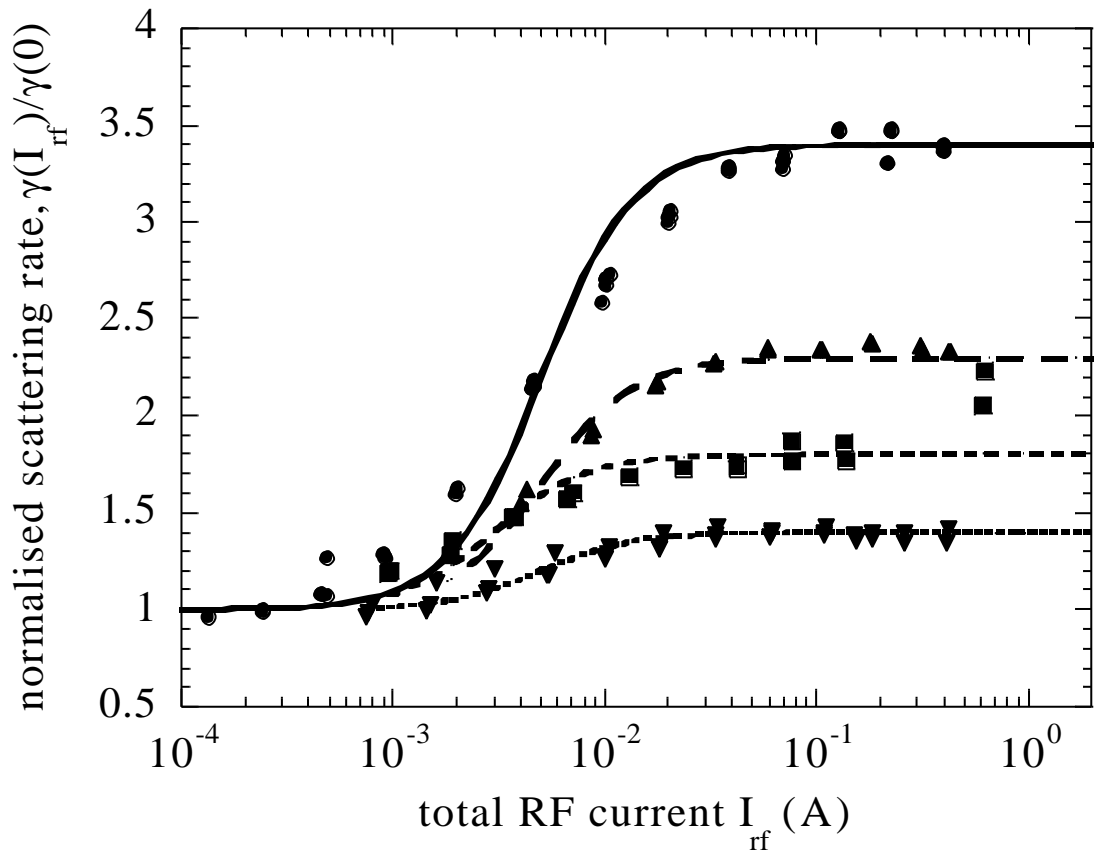

Figure 7.

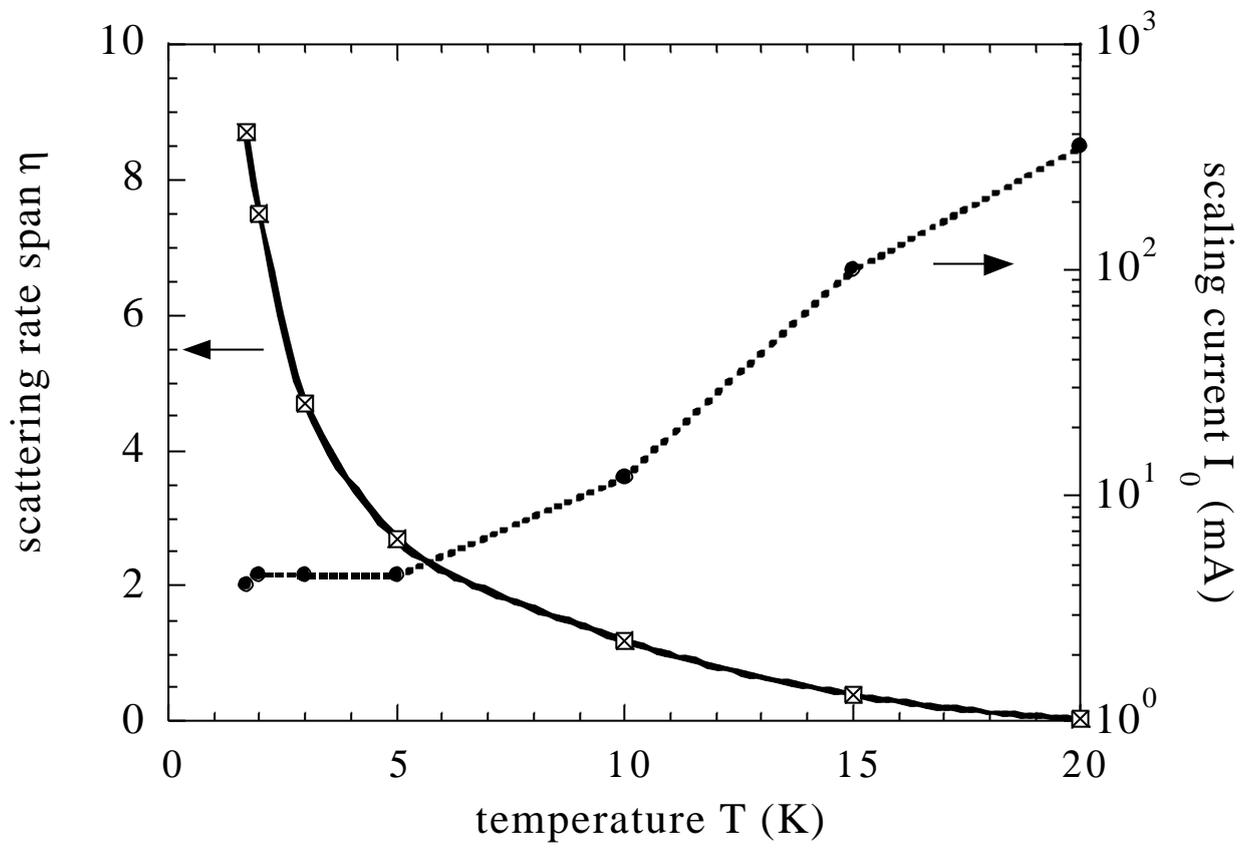

Figure 8.

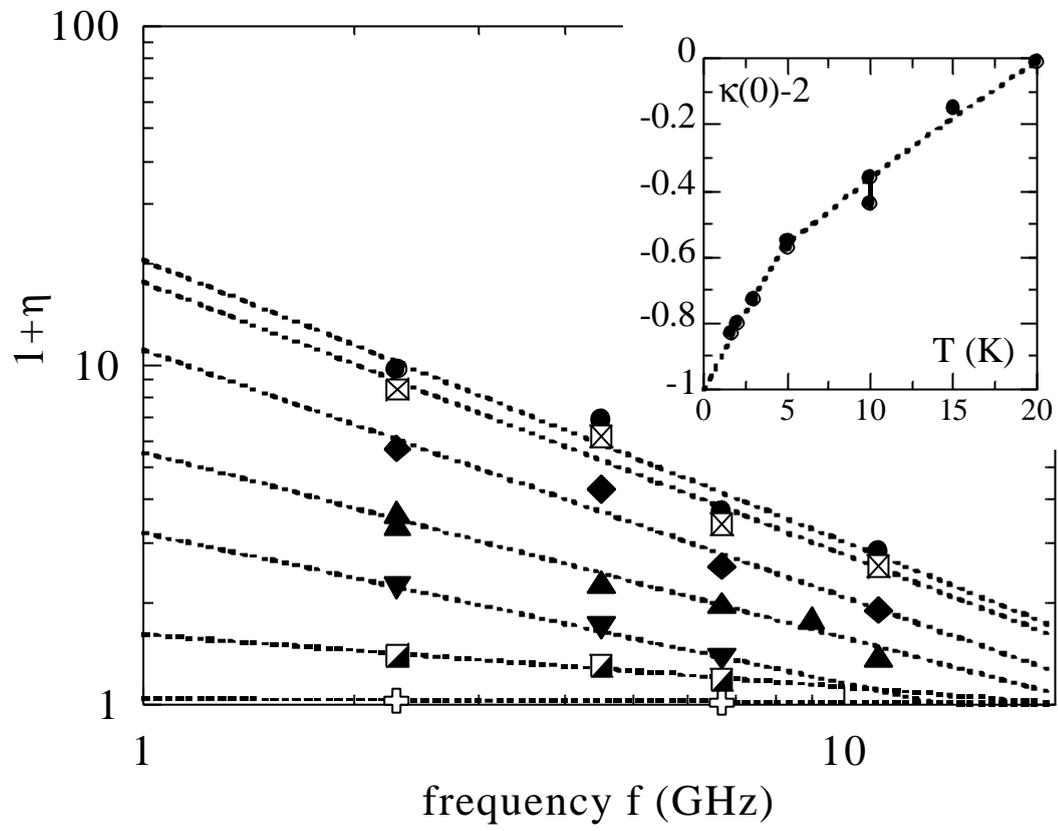

Figure 9.